\documentclass[showpacs,floatfix,a4paper,prb,twocolumn]{revtex4}
\usepackage{amssymb}
\usepackage{amsfonts}
\usepackage{amsmath}
\usepackage{graphicx}
\usepackage{latexsym,epsfig,bm,times,psfrag,subfigure}
\usepackage{bm,times}
\usepackage{dcolumn}

\begin{document}

\title{Equilibration of integer quantum Hall edge states }
\author{D.\ L.\ Kovrizhin and J.\ T.\ Chalker}
\affiliation{Theoretical Physics, Oxford University, 1, Keble Road, Oxford, OX1 3NP,
United Kingdom}
\date{\today}

\begin{abstract}
We study equilibration of quantum Hall edge states at integer filling
factors, motivated by experiments involving point contacts at finite bias.
Idealising the experimental situation and extending the notion of a quantum
quench, we consider time evolution from an initial non-equilibrium state in
a translationally invariant system. We show that electron interactions bring
the system into a steady state at long times. Strikingly, this state is not
a thermal one: its properties depend on the full functional form of the
initial electron distribution, and not simply on the initial energy density.
Further, we demonstrate that measurements of the tunneling density of states
at long times can yield either an over-estimate or an under-estimate of the
energy density, depending on details of the analysis, and discuss this finding in
connection with an apparent energy loss observed experimentally. More
specifically, we treat several separate cases: for filling factor $\nu{\ =}1$
we discuss relaxation due to finite-range or Coulomb interactions between
electrons in the same channel, and for filling factor $\nu{=}2$ we examine
relaxation due to contact interactions between electrons in different
channels. In both instances we calculate analytically the long-time
asymptotics of the single-particle correlation function. These results are
supported by an exact solution at arbitrary time for the problem of
relaxation at $\nu =2$ from an initial state in which the two channels have
electron distributions that are both thermal but with unequal temperatures,
for which we also examine the tunneling density of states. 
\end{abstract}

\maketitle

\section{Introduction}

Recent experiments have stimulated renewed interest in the process of
equilibration for an isolated quantum system. Some of the most prominent of
these studies have involved time evolution in cold atomic gases, after a
sudden change in system parameters.\cite{kinoshita,schmiedmayer} Other
measurements, remarkably, are able to probe coherent quantum dynamics in
conventional condensed matter systems far from equilibrium.\cite%
{heiblum2,grangerheat,pierre,pierrenu2,altimiras,cavalleri} Theoretical work
has included studies of general features of dynamics after a quantum quench,%
\cite{calabrese} and of the approach to thermal equilibrium,\cite{olshanii}
as well as detailed investigations of individual model systems.\cite%
{mirlin,numerics}

A notable result from experiments on cold atomic gases is the absence of
relaxation in some cases, despite there being strong interactions between
particles.\cite{kinoshita} This has been attributed to the fact that the
system concerned is, to a good approximation, described by the Hamiltonian
for an integrable model.

In the condensed matter setting, a series of recent investigations have
examined aspects of the physics of quantum Hall edge states far from
equilibrium. These include: the discovery of interaction effects in
electronic Mach-Zehnder interferometers;\cite{heiblum2} the detection of
thermal transport by quantum Hall edge states;\cite{grangerheat} and the
measurement\cite{pierre} of the energy distribution of electrons in an edge
state driven out of equilibrium by passage through a quantum point contact
(QPC) at finite bias voltage, and the observation\cite{pierrenu2} and control%
\cite{altimiras} of the relaxation of this distribution.

The theoretical work we describe in this paper is motivated in part by the
last of these sets of experiments. In outline, the experiments\cite%
{pierre,pierrenu2,altimiras} involve bringing together, at a QPC, integer
quantum Hall edge states that originate from contacts at different
potentials. Tunneling between edges within the QPC generates a
non-equilibrium electron energy distribution downstream from the contact.
This distribution is measured\cite{pierre} using a quantum dot as a
spectrometer, as a function of propagation distance.\cite{pierrenu2} It
relaxes to a stationary form, which is close to thermal but with an
effective temperature higher than that of the sample as a whole. The fact
that this distribution propagates further with little change suggests that
electrons in edge states are weakly coupled to other degrees of freedom.

These observations raise some important theoretical questions. A central one
stems from the fact that the usual, chiral Luttinger liquid model for quantum Hall edge states\cite%
{wen} is integrable, being simply quadratic in the edge magnetoplasmon
coordinates. Within this description, the number of excitation quanta in
each collective mode is conserved under time evolution. In this context, it
is natural to ask how relaxation takes place in edge states, and what steady
state is reached at long times. %

The main difficulty in addressing these issues arises because interactions
are simplest to handle using bosonization, while a QPC is most naturally
represented using fermionic variables. The two existing theoretical
discussions\cite{buttiker,degiovanni} of edge state relaxation aim at a
faithful, but necessarily approximate, treatment of the experimental
arrangement, including the QPC. Our objective in this paper is to explore a
complementary direction. We study interaction effects on time evolution in a
translationally invariant system, omitting the QPC altogether from our model
Hamiltonian. In its place, we choose the initial state of the system to have
an electron distribution in momentum like that generated when
non-interacting fermions pass through a tunneling contact at finite bias
voltage. Our problem therefore has the features of a quantum quench,\cite%
{calabrese} except that the initial state is not chosen to be the ground
state of a simple Hamiltonian. In our treatment we expect evolution as a
function of time to correspond qualitatively to evolution in the
experimental system as a function of distance from the QPC, with time and
distance in the two pictures related by a characteristic edge state
velocity. While this correspondence is not precise, our exact treatment of
interactions brings advantages. In particular, we are able to show that the
one-electron correlation function reaches a stationary form in the long-time
limit that is \emph{not thermal}. It would probably be hard to know whether
this conclusion was reliable if it were the result of an approximate
calculation. Our calculations draw on theoretical ideas developed in recent
discussions\cite{chalker07,Levkivskyi,kovrizhin_mzprb} of non-equilibrium
effects in Mach-Zehnder interferometers.

The remainder of the paper is organised as follows. In Section \ref%
{sec:discussion} we give an overview of the results and discuss the relation
of our approach to experiments and to previous theory. 
In Section \ref{sec:model} we introduce the
model and notation we use. In Section \ref{sec:nu=1} we show how plasmon
dispersion arising from finite range or Coulomb interactions in an edge at $%
\nu=1$ generates a stationary electron distribution at long times, for an
initial state that has no many-particle correlations but it is otherwise
general. We then specialise this treatment to an initial state in which the
electron distribution in momentum consists of a double step. In Section \ref%
{sec:nu=2short} we treat a system at $\nu=2$ with contact interactions
between electrons in different channels. We obtain an exact solution for the
time evolution with edges initially at different temperatures, and study the
long-time distribution reached for a general initial state. Finally, in
Section \ref{different-t} we use this solved example to discuss the
tunneling density of states and implications for experimental estimates of
the energy density in the steady state. 

\section{Overview of results}

\label{sec:discussion}

In the experiments\cite{pierrenu2} we are concerned with, edge channels of a
quantum Hall system at filling factor $\nu =2$ are driven out of equilibrium
by applying bias voltage $V$ to a QPC. Edges that originate from contacts
with different potentials meet at the QPC where, depending on its width,
either the inner or the outer channels of the two edges are coupled by
quantum tunneling. As a result, immediately after the QPC, the electrons in
these channels have a non-equilibrium energy distribution. For example, in a
non-interacting system at zero temperature, this distribution has two steps
with energy separation $eV$, as shown in Fig.~\ref{fig:1}. The energy
distribution of electrons is expected to evolve as a function of distance
downstream from the QPC, because of inelastic scattering induced by
interactions. This distribution is measured in the outer channel at
successive distances from the QPC, using a quantum dot with a single active
electronic level. It is found that the double-step distribution generated by
the QPC equilibrates after a propagation length of order $10\ \mu \mathrm{m}$%
. The observed equilibrium distribution is similar to a Fermi function, with
an effective temperature close to that set by the average electron energy density
in the channels just after the QPC.

Two different approaches have been used in previous work to model 
these experiments. One\cite{buttiker} is formulated in terms of electron operators,
using a Boltzmann-like equation to study the evolution of the electron distribution with 
distance as a consequence of electron-electron interactions. This method
has the advantage that the initial electron distribution can be calculated 
simply by using a single-particle treatment of the QPC. Its limitation 
stems from the fact that the integrability of the usual, chiral Luttinger liquid model for edge states 
is presumably not respected. The other approach\cite{degiovanni} is formulated in terms
of edge magnetoplasmons. This allows a detailed treatment of interactions,
at the cost of a phenomenological treatment of the QPC, in which the plasmon 
distribution just after the QPC is estimated from shot noise in the charge.

We evade some of these difficulties by treating 
relaxation of an initial distribution in a translation-invariant edge. The relaxation is
caused by scattering of electrons between different single-particle momentum
eigenstates. By contrast, the collective plasmon modes of the system do not
relax, because within standard approximations the bosonized Hamiltonian is
quadratic. Instead, using the language of plasmons, relaxation occurs
because in the interacting system different modes have different group
velocities. To expand on this viewpoint, we use it to estimate the
relaxation time for an initial state in which electrons occupy
single-particle levels independently, with a specified energy distribution.
This state can be viewed in terms of particle and hole excitations above a
Fermi sea. It is characterised by the mean separation $L_{\mathrm{init}}$
between these excitations. For an electron energy distribution having two
steps at energy separation $eV$, with edge velocity $v$, we have $L_{\mathrm{%
init}}\sim \hbar v/eV$. Because of plasmon dispersion, the width $\ell$ of
particle and hole wavepackets broadens with time $t$: the relaxation time $%
\tau$ is the time scale at which their width is comparable to their
separation.

The simplest application of these ideas is provided by a system at filling
factor $\nu =2$ with contact interactions between electrons in different
channels. Plasmon dispersion in this situation is characterised by
velocities $v_{-}$ and $v_{+}$ for slow modes and fast modes, as reviewed in
Section \ref{sec:nu=2short}. Then the width of the wavepacket grows as
$\ell \sim (v_{+}-v_{-})t$ and the relaxation time is 
\begin{equation}\label{tau}
\tau =\frac{v_{+}+v_{-}}{v_{+}-v_{-}}\frac{\hbar }{eV}\,.
\end{equation}%
Note that this timescale and the corresponding lengthscale (similar to one
identified\cite{footnote} in Ref.~\onlinecite{degiovanni}), vary inversely with bias voltage 
$V$. It would be interesting to search for such a dependence in experiment: this might be done
from the data in Fig.~2 of the Supplementary Material to Ref.~\onlinecite{pierrenu2}, by extracting the decay length 
of $T_{\rm exc}$ and analysing its dependence on $\delta V_{\rm D}$.

As a second example, consider a single edge channel at filling factor $\nu
=1 $ with finite range interactions. The plasmon frequency $\omega _{q}$
varies with wavevector $q$, having the small-$q$ expansion 
\begin{equation}
\omega _{q}=vq-(v/b)(bq)^{3}/3\ldots ,  \label{nldisp}
\end{equation}%
where $b>0$ is a length that characterises the range and strength of
interactions. In place of $v_{+}-v_{-}$ we take the difference between the
group velocity $\partial _{q}\omega _{q}$ at $q=0$ and at $q=\ell ^{-1}$,
which is $v(b/\ell )^{2}$. Setting $\ell=vt(b/\ell)^{2}$ yields $%
\ell=b(vt/b)^{1/3}$ and from $\ell=L_{\mathrm{init}}$ we obtain 
\begin{equation}
\tau =\left( \frac{v}{b}\right) ^{2}\left( \frac{\hbar }{eV}\right) ^{3}\,.
\end{equation}%
Thus at $\nu =1$ with finite range interactions, we find a timescale and
lengthscale that vary with bias voltage as $V^{-3}$, in striking contrast to
behaviour at $\nu =2$. Again, it would be of interest to test this in
experiment.

For unscreened Coulomb interactions the plasmon dispersion is 
\begin{equation}
\omega_q=q[v+u\ln (1/bq)]  \label{coulombdisp}
\end{equation}%
and the width of the electron wave-packet grows with time as $\ell =ut$,
giving 
\begin{equation}
\tau = \frac{\hbar u}{evV}\,.
\end{equation}
Similar scaling relations have been discussed in Ref.\onlinecite{chalker07}
in the context of the decoherence in electronic Mach-Zehnder interferometers.

We show in the remainder of this paper that at times much longer than $\tau$%
, the equal time, one-electron correlation function reaches a stationary
form. For $\nu =1$ and an initial state in which single particle levels are
occupied independently, we determine this form exactly as a functional of
the initial distribution. Its short-distance behaviour is fixed by the
energy density in the nonequilibrium state, and characterised by an
effective temperature. Its behaviour at long distances, however, depends on
the full functional form of the initial distribution. We present detailed
results for the case of a double-step initial distribution, as produced at a
QPC. For this example we also consider the electron momentum distribution at
long times, which shows power-law behavior at small momenta, with an
exponent which depends on tunneling probability of the QPC. At large momenta
the distribution is close to thermal with an effective temperature given by
the excess energy.

The electron distribution is probed in the experiments of Refs.~%
\onlinecite{pierre,pierrenu2,altimiras, otsuka} by measurement of the tunneling
current through a quantum dot that has a single active energy level weakly
coupled to the edge state. Without interactions, this current is
proportional to the occupation probability of edge-state orbitals at the
energy level of the dot, and so yields a straightforward determination of
the electron distribution in energy. With interactions, however, the
situation is more complicated, since matrix elements for tunneling between
the edge and the dot depend on the many-body state of the quantum Hall edge.
We examine the consequences of interaction effects for the interpretation of
tunneling current measurements in Section \ref{different-t}, focussing in
particular on estimates of the energy density. We find that the energy
density can be over-estimated or under-estimated, depending on how such
measurements are analysed. 

\section{Model and observables}

\label{sec:model}

In the following we study edge states in the integer quantum Hall regime at
filling factors $\nu =1$ and $\nu =2$. The microscopic Hamiltonian $\hat{H}$
is a sum 
\begin{equation}
\hat{H}=\hat{H}_{kin}+\hat{H}_{int}  \label{hamedge}
\end{equation}%
of contributions from kinetic energy $\hat{H}_{kin}$ and interaction energy $%
\hat{H}_{int}$. For $n_{c}$ chiral copropagating electron channels, with
channel index $\eta $ and Fermi-velocities $v_{\eta }$, we have 
\begin{equation}
\hat{H}_{kin}=-i\hbar \sum\limits_{\eta =1}^{n_{c}}v_{\eta }\int_{-L/2}^{L/2}%
\hat{\psi}_{\eta }^{+}(x)\partial _{x}\hat{\psi}_{\eta }(x)dx\,.
\label{hkin}
\end{equation}%
Here we impose periodic boundary conditions on a system of length $L$ so
that allowed wavevectors are $k=2\pi n/L$, with integer $n$. Fermionic
creation and annihilation operators $\hat{c}_{k\eta }^{+}$ and $\hat{c}%
_{k\eta }$ for an electron with wavevector $k$ on the edge $\eta$ obey
standard anticommutation relations $\{\hat{c}_{k\eta },\hat{c}_{p\eta
^{\prime }}^{+}\}=\delta _{kp}\delta _{\eta \eta ^{\prime }}$. The field
operator $\hat{\psi}_{\eta }(x)$, which annihilates an electron at position $%
x$ on the channel $\eta $, is 
\begin{equation}
\hat{\psi}_{\eta }(x)=\frac{1}{\sqrt{L}}\sum_{k=-\infty }^{\infty }\hat{c}%
_{k\eta }e^{ikx}.  \label{psi}
\end{equation}%
We consider translationally-invariant interactions described by the
potential $U_{\eta \eta ^{\prime }}(x-x^{\prime })$, so that 
\begin{equation}
\hat{H}_{int}=\frac{1}{2}\int_{-L/2}^{L/2}\int_{-L/2}^{L/2}dxdx^{\prime }\
U_{\eta \eta ^{\prime }}(x-x^{\prime })\hat{\rho}_{\eta }(x)\hat{\rho}_{\eta
^{\prime }}(x^{\prime }).
\end{equation}%
We discuss two different cases: a single edge channel at $\nu =1$, with
finite range interactions; and a two-channel system at $\nu =2$, with
contact interactions between electrons in different channels. Contact
interactions between electrons in the same channel can be absorbed into a
renormalization of Fermi velocities.

Our first objective is to obtain the equal time single-particle correlation
function. Using a time argument to indicate operators $\hat{\psi}_{\eta
}(x,t)=e^{i\hat{H}t}\hat{\psi}_{\eta }(x)e^{-i\hat{H}t}$ in their Heisenberg
representation, we write this quantity at time $t$ as 
\begin{equation}
G_{\eta }(x-x^{\prime },t)=\langle \hat{\psi}_{\eta }^{+}(x,t)\hat{\psi}%
_{\eta }(x^{\prime },t)\rangle ,  \label{gfun1}
\end{equation}%
where the average is calculated with respect to the initial state, yet to be
specified. The electron momentum distribution in a translationally invariant
system is then 
\begin{equation}
n_{\eta }(k,t)=\int_{-L/2}^{L/2}dx\ e^{ikx}G_{\eta }(x,t).  \label{mom_distr}
\end{equation}

In thermal equilibrium at inverse temperature $\beta $ the single-particle
correlation function has at $\nu=1$ the form 
\begin{equation}
G(x,t)=i\left\{ {2\beta \hbar v\sinh [{\pi }(x-vt+ia)/\beta \hbar v]}%
\right\} ^{-1}\   \label{freeGF}
\end{equation}%
with a lengthscale $\beta \hbar v$, timescale $\beta\hbar$ and a
short-distance cut-off $a$. 

The experiments of Refs.~\onlinecite{pierre,pierrenu2,altimiras} measure the
current through a quantum dot (QD) weakly coupled to one edge channel.
Assuming sequential incoherent tunneling through a QD with a single energy
level $E$, perturbation theory in the tunneling amplitude $t_{D}$ gives for
this current the expression\cite{bruus} 
\begin{equation}
\frac{e|t_{D}|^{2}}{\hbar }\nu _{\eta }(E),  \label{IQD2}
\end{equation}%
where the tunneling density of states is at long times%
\begin{equation}\label{DOS1}
\nu _{\eta }(E)=\lim_{t \to \infty}\hbar^{-1}\int_{-\infty }^{\infty }d\tau \ \langle \hat{%
\psi}^{+}(0,t+\tau )\hat{\psi}(0,t)\rangle e^{-iE\tau/\hbar }.
\end{equation}%

\section{Equilibration at $\protect\nu{=}1$}

\label{sec:nu=1}

To compute the time evolution of operators under $\hat{H}$ we use
bosonization, \cite{vonDelft} employing the same notation as in section IVC
of~Ref.~\onlinecite{kovrizhin_mzprb}. Consider a system at $\nu =1$. The
Hamiltonian in bosonic form is 
\begin{equation}
\hat{H}=\sum_{q>0}\hbar \omega _{q}\hat{b}_{q}^{+}\hat{b}_{q}  \label{Hnl}
\end{equation}%
where $\hat{b}_{q}^{+}$ is an operator creating a plasmon with wavevector $q$
and energy $\hbar \omega _{q}$, and we omit the channel label $\eta$. Let
the Fourier transform of the interaction potential be 
\begin{equation}
u(q)=\int_{-\infty }^{\infty }dx\ e^{-iqx}U(x)\,.
\end{equation} 
Then the boson dispersion relation is 
\begin{equation*}
\omega _{q}=vq[1+u(q)/2\pi \hbar v]\,.
\end{equation*}

In this section we express the correlation function, Eq.~(\ref{gfun1}), in
terms of the electron density operator $\hat{\rho}(x) = \hat{\psi}^+(x)\hat{\psi}(x)$
written in the Schr\"odinger representation. The result, Eq.~(\ref{GFK1}),
involves an average of $\hat{\rho}(y)$ with a kernel, denoted below by $%
K(x,t;y)$. Quantum fluctuations of this average are Gaussian at long times,
because the range of the kernel becomes much larger than the correlation
length in the initial state. For this reason, the correlation function in
the long time limit can be expressed as a functional of the second moment of
the density in the initial state. More formally, one can write a cumulant
expansion for the correlation function, in which the second order cumulant
is time independent and higher order contributions decay to zero at long
times if the plasmon dispersion $\omega _{q}$ is not exactly linear in $q$.

\subsection{Correlations at $\protect\nu=1$ in terms of initial density
operator}

\label{correlations-via-density}

Introduction of the bosonic field 
\begin{equation}
\hat{\phi}\left( x,t\right) =-\sum_{q>0}(2\pi /qL)^{1/2}(\hat{b}%
_{q}e^{iqx-i\omega _{q}t}+\mathrm{h.c})e^{-qa/2}  \notag
\end{equation}%
allows the electron correlation function to be written as 
\begin{equation}
G(x,t)=(2\pi a)^{-1}\langle e^{i\hat{\phi}(x,t)}e^{-i\hat{\phi}(0,t)}\rangle
\,.  \label{GF1}
\end{equation}%
In order to evaluate this average, we express the boson operators $\hat{b}%
_{q}$ and $\hat{b}_{q}^{+}$ in terms of $\hat{\rho}(x)$. From the standard
expression 
\begin{equation*}
\hat{\rho}(y)=\frac{1}{2\pi }\sum_{q>0}(2\pi /qL)^{1/2}[iq\hat{b}%
_{q}e^{iqy}-iq\hat{b}_{q}^{+}e^{-iqy}],
\end{equation*}%
we have for $k>0$ 
\begin{equation*}
\int_{-L/2}^{L/2}\hat{\rho}(y)e^{-iky}dy=\frac{L}{2\pi }\left( \frac{2\pi }{%
kL}\right) ^{1/2}ik\hat{b}_{k}.
\end{equation*}%
Thus the difference in bosonic fields 
\begin{equation*}
\hat{\phi}(x,t)-\hat{\phi}(0,t)=\int_{-L/2}^{L/2}K(x,t;y)\hat{\rho}(y)dy,
\end{equation*}%
is given by a spatial average of the electron density operator weighted with
the kernel 
\begin{equation}
K\left( x,t;y\right) =-i\sum_{q\neq 0}\frac{2\pi }{qL}(e^{-iqx}-1)e^{i\omega
_{q}t+iqy}.  \label{kernel_lr}
\end{equation}%
This yields an expression for the correlation function in terms of fermionic
operators 
\begin{equation}
G(x,t)=(2\pi a)^{-1}e^{I(x,a)/2}\langle e^{i\int dy\ K(x,t;y)\hat{\rho}%
\left( y\right) }\rangle \,,  \label{GFK1}
\end{equation}%
where%
\begin{equation}
I(x,a)\equiv \lbrack \hat{\phi}(x,t),\hat{\phi}(0,t)]=\ln [(a+ix)/(a-ix)].
\label{Iacomm}
\end{equation}

\subsubsection{Long-time asymptotics of the kernel}

\label{sec:kernel}

Considered as a function of $y$, the kernel $K\left( x,t;y\right) $ at time $%
t=0$ is a step of width $|x|$ with height $-2\pi $. In the noninteracting
system for $t>0$, this step propagates at velocity $v$ without change in
shape. With interactions, the relaxation time of $\tau$ introduced in Section II is
the timescale for the step to change shape. We are interested in behavior of $K$ at long times in presence of
interactions. Consider first finite-range interactions with the plasmon
dispersion at small wavevector given by Eq. (\ref{nldisp}). We separate out
the average motion by writing $y=x/2-vt+\lambda $. This gives for the phase
appearing in Eq.~(\ref{kernel_lr}) 
\begin{equation}
q(y-x/2)+\omega _{q}t=q\lambda -(vt/b)(bq)^{3}/3+\ldots
\end{equation}%
at small $q$. Defining a lengthscale $l_{S}=b(vt/b)^{1/3}$ as in Section \ref%
{sec:discussion}, the kernel at long times has the form 
\begin{equation}
K(x,t,y)=-2\int_{-\infty }^{\infty }\frac{dQ}{Q}\sin \frac{Qx}{2l_{S}}\cos
[Q(\lambda /l_{S})+Q^{3}/3]\,.  \label{KS}
\end{equation}%
Explicit evaluation for $|x|\ll l_{S}$ yields 
\begin{equation}
K(x,t,y)=-2\pi (x/l_{S})\mathrm{Ai}[\lambda /l_{S}]\,.
\end{equation}%
Thus the range of the kernel grows with time as $t^{1/3}$ and its amplitude
decays as $t^{-1/3}.$

We can obtain asymptotics of the kernel for a system with Coulomb
interactions in a similar way. In this case the plasmon dispersion is given
by Eq.~(\ref{coulombdisp}). Writing $y=x/2-vt-ut\ln (ut/b)+\xi $, the phase
appearing in Eq.~(\ref{kernel_lr}) is 
\begin{equation}
q(y-x/2)+\omega _{q}t=q\lambda +qut\ln (1/qut)\,.
\end{equation}%
Defining $l_{C}=ut$, we have%
\begin{equation}
K(x,t,y)=-2\int_{-\infty }^{\infty }\frac{dQ}{Q}\sin \frac{Qx}{2l_{C}}\cos
(Q\lambda /l_{C}-Q\ln |Q|)]\,,
\end{equation}%
Hence, with Coulomb interactions at long time, the range of the kernel grows
as $t$ and its amplitude decays as $t^{-1}$.

\subsubsection{Cumulant expansion}

To evaluate the average in Eq.~(\ref{GFK1}) we write it in terms of
cumulants, as 
\begin{multline}  \label{cumulants}
\langle e^{i\int K\left( x,t;y\right) \hat{\rho}\left( y\right) dy}\rangle
=\exp \bigg[\sum_{n}\frac{i^{n}}{n!}\int_{-L/2}^{L/2}dy\ dz_{1}\ldots
dz_{n-1} \\
\times F_{n}(x,t,y;z_{1}\ldots z_{n-1})C_{n}\left( z_{1}\ldots
z_{n-1}\right) \bigg] \,,
\end{multline}
where $n$-th order cumulant of the electron density $\hat{\rho}(y)$ is 
\begin{equation}
C_{n}\left( z_{1},...z_{n-1}\right) =\langle \rho \left( y\right) \rho
\left( y+z_{1}\right) \ldots \rho \left( y+z_{n-1}\right) \rangle  \label{CN}
\end{equation}%

and the $n$-th order convolution of kernels is 
\begin{multline*}
F_{n}(x,t;z_{1}\ldots z_{n-1})=\frac{1}{2\pi }\int_{-L/2}^{L/2}dy\ K(x,t;y)
\\
\times K(x,t;y+z_{1})\ldots K\left( x,t;y+z_{n-1}\right) .
\end{multline*}%
Consider the second order contribution. This is 
\begin{multline}
F_{2}(x,t;z)=\frac{1}{2\pi }\int_{-L/2}^{L/2}dy\ K(x,t;y)K(x,t;y+z)= \\
\frac{1}{2\pi }(-i)^{2}\left( 2\pi /L\right) ^{2}\sum_{q,k\neq 0}\frac{1}{qk}%
\left( e^{-iqx}-1\right) \left( e^{-ikx}-1\right) \\
\times \int_{-L/2}^{L/2}dy\ e^{i(q+k)y}e^{ikz}e^{i(\omega _{q}+\omega
_{k})t}\,.
\end{multline}
The result, independent of both time and interactions, is 
\begin{equation}
F_{2}(x,z)=\pi (|x+z|+|x-z|-2|z|)\,.  \label{F2}
\end{equation}%
It is a triangle as a function of $z,$ which is non-zero in the region $%
(-x,x)$ and has amplitude $F_{2}(x,0,0)=2\pi \left\vert x\right\vert $.

At third order we have 
\begin{multline*}
F_{3}(x,t;z_{1},z_{2})=-\left( -i\right) ^{3}\int_{-\infty }^{\infty }\frac{
dq}{q}\int_{-\infty }^{\infty }\frac{dp}{p}\frac{1}{p+q}\times \\
\times (e^{-iqx}-1)\left( e^{-ipx}-1\right) (e^{i(p+q)x}-1) \\
\times e^{i(\omega _{p}+\omega _{q}-\omega
_{p+q})t}e^{ipz_{1}}e^{-i(p+q)z_{2}}.
\end{multline*}
Time enters the integrand of this expression through the factor $e^{i(\omega
_{p}+\omega _{q}-\omega_{p+q})t}$. Without interactions $\omega _{q}+\omega
_{p}-\omega _{p+q}=0$, but in an interacting system this factor oscillates
rapidly at long times, as a function of $p$ and $q$. For large $t$ the
integrals are therefore small and are dominated by low-frequency
contributions. In this regime we can use the expansion $\omega _{q}+\omega
_{p}-\omega _{p+q}\propto q^{3}+p^{3}-[p+q]^{3}$ and remove the time
dependence from the exponent by rescaling variables, with $q=Q/l_S$, $%
p=P/l_S $, as in the discussion of the kernel, and $x=Xl_S$, $%
z_{1,2}=Z_{1,2}l_S\ $. In this way we find $F_{3}=t^{-1/3}\times $ (a
time-independent function of $X,Z_{1},Z_{2}).$ Because the functions $%
C_{n}\left( z_{1}\ldots z_{n-1}\right) $ decay at large $z$ we can use in
Eq.~(\ref{cumulants}) the function $F_{3}$ evaluated at $Z_1=Z_2=0$,
obtaining a contribution that decays as $t^{-1/3}$. An equivalent result
holds for all higher order cumulants. Similar arguments also apply in the
case of Coulomb interactions, but with third and higher order terms decaying
in time as $t^{-1}$. 

The significance of these results is that contributions from all cumulants
beyond the second one decay with time and may be neglected in the long-time
limit. Cancellation of the high order cumulants is known for thermal states
as the Dzyaloshinskii-Larkin theorem,\cite{dzyaloshinskii,mirlin} but is not in
general valid for nonequilibrium states. We have thus shown that a similar
result holds in the long-time limit.

In summary, we obtain for the correlation function in the long time limit
the expression 
\begin{multline}
\mathcal{G}(x)=\lim_{t\rightarrow \infty }G(x,t)=(2\pi a)^{-1}e^{\frac{1}{2}
I(a)}\times  \label{GLT2} \\
\exp \left[ -\pi \int_{-\infty }^{\infty }dz\ F_{2}(x,z)\langle \hat{\rho}%
(x) \hat{\rho}(x+z)\rangle \right] \,.
\end{multline}
This is one of the main results of this paper. In order to use it, we need
to calculate $\langle \hat{\rho}(y)\hat{\rho}(y+z)\rangle $ for the initial
state, which we do below.

\subsection{Long-time correlations}

\label{steady-state-nu=1}

As a check and because we shall need to refer to the results, we start by
considering averages in the ground state, denoted by $\langle \ldots \rangle
_{0}$. We require $\langle \hat{\rho}(0)\hat{\rho}(z)\rangle _{0}$ and find 
\begin{equation*}
\langle \hat{\rho}(0)\hat{\rho}(z)\rangle _{0}=\frac{1}{(2\pi )^{2}}\frac{1}{%
(a+iz)^{2}}.
\end{equation*}%
Integrating this with the function $F_{2}(x,z)$ [Eq. (\ref{F2})] we have 
\begin{multline}
\int_{-\infty }^{\infty }dz\ F_{2}(x,z)\langle \hat{\rho}(x)%
\hat{\rho}(x+z)\rangle _{0}  \label{FRHO} \\
=-\frac{1}{2\pi}[\ln a^{2}-\ln (x^{2}+a^{2})].
\end{multline}%
Substituting Eqns.~(\ref{FRHO}) and (\ref{Iacomm}) into Eq.~(\ref{GLT2}) we
obtain for the ground state correlation function 
\begin{multline*}
G(x,0)=(2\pi a)^{-1}\exp \{\frac{1}{2}[\ln a^{2}-\ln (x^{2}+a^{2})]\} \\
\times \lbrack (a+ix)/(a-ix)]^{1/2}=\frac{i}{2\pi }\frac{1}{x+ia}\,,
\end{multline*}
in agreement with Eq.~(\ref{freeGF}), evaluated for $\beta \to \infty$.

Now consider an initial state in which single-particle momentum eigenstates
are occupied independently with probability $n(k)$ as a function of
wavevector $k$. We separate this probability into its value $n_{0}(k)$ for a
filled Fermi sea with Fermi wavevector zero, and a deviation $\delta n(k)$,
writing $n(k)=n_{0}(k)+\delta n(k)$. Without loss of generality, we take 
\begin{equation}
\int_{-\infty }^{\infty }\delta n(k)\ dk=0.  \label{eqnqz}
\end{equation}%
In this state we have 
\begin{eqnarray}
C_{2}(z) &=&\frac{1}{L^{2}}\sum_{k_1k_2k_3} \langle \hat{c}_{k_{1}}^{+}\hat{c}_{k_{2}}%
\hat{c}_{k_{3}}^{+}\hat{c}_{k_{1}+k_3-k_2}\rangle
e^{i(k_{1}-k_{2})z}  \notag \\
&=&\langle \hat{\rho}(0)\hat{\rho}(z)\rangle
_{0}+C_{2}^{(1)}(z)+C_{2}^{(2)}(z)\,
\end{eqnarray}%
where we have introduced 
\begin{equation}
C_{2}^{(1)}(z)=\frac{1}{2\pi ^{2}z}\int_{-\infty }^{\infty }dq\,\delta
n(q)\sin qz\,.
\end{equation}%
and 
\begin{equation}
C_{2}^{(2)}(z)=-\left\vert \frac{1}{2\pi }\int dq\ \delta
n(q)e^{-iqz}\right\vert ^{2}.
\end{equation}%
So in the long time limit we have 
\begin{equation}
\mathcal{G}(x)=\frac{i}{2\pi }\frac{1}{x+ia}\exp [-f\left( x\right) ]
\label{GFX}
\end{equation}%
with 
\begin{equation}
f(x)=\pi \int_{-\infty }^{\infty }dz\
F_{2}(x,z)[C_{2}^{(1)}(z)+C_{2}^{(2)}(z)]\,.  \label{FX}
\end{equation}%
This represents a second key result of our work, and we next analyse the
behaviour of $f(x)$.

\subsubsection{Short-distance asymptotics}

Consider $f(x)$ for $x$ much smaller than the quasiparticle spacing in the
initial state. 
Since $F(x,z)$ is non-zero only for $|z|<|x|$ we require the small $z$
behavior of $C_{2}^{(1)}(z)$ and $C_{2}^{(2)}(z)$. Both can be expressed in
terms of the energy density in the initial state, characterised by an
effective temperature $T^{\ast }$. This is defined following the relation 
\begin{equation}
\frac{\hbar v_{F}}{2\pi }\int \,\delta n(q)\,q\,\mathrm{d}q=\frac{\pi }{12}%
(k_{\mathrm{B}}T)^{2}  \label{EDST}
\end{equation}%
between energy density and temperature $T$ for a thermal state. We have $%
C_{2}^{(1)}(z)=(k_{\mathrm{B}}T^{\ast }/{\hbar v_{F}})^{2}/12+\mathcal{O}%
(z^{2})$ and $C_{2}^{(2)}(z)=-[\pi zC_{2}^{(1)}]^{2}$. %
In this way we find 
\begin{equation}
f(x)=\frac{\pi ^{2}}{6}\left( \frac{k_{\mathrm{B}}T^{\ast }}{\hbar v_{F}}%
\right) ^{2}x^{2}+\mathcal{O}(x^{4})\,.  \label{Fsmallx}
\end{equation}%
\textit{In summary, the short-distance behaviour of the steady state is
identical to that in thermal equilibrium at a temperature fixed by the
initial energy density.}

\subsubsection{Long-distance asymptotics}

At large $x$, the function $F_{2}(x,z)$ has slow dependence on $z$, but $%
C_{2}^{(1)}(z)$ and $C_{2}^{(2)}(z)$ fall to zero for $|z|$ much greater
than the quasiparticle spacing in the initial state. We can therefore make
the replacement $F_{2}(x,z)=2\pi |x|$ in Eq.~(\ref{FX}), obtaining 
\begin{equation}
f(x)={\pi |x|}\int_{-\infty }^{\infty }dq\ \left[ \mathrm{sgn}(q)\ \delta
n(q)-|\delta n(q)|^{2}\right] +\mathcal{O}(|x|^{0})\,.  \label{Flargex}
\end{equation}%
\textit{The long-distance decay of the correlation function is therefore not in
general determined solely by the energy density, but instead depends on the
entire momentum distribution in the initial state.} In the following
subsection we examine the consequences of these general results for a
specific choice of initial distribution.

\subsubsection{Equilibration from double-step initial momentum distribution}

\label{sec:nu=1b}

\begin{figure}[b]
\epsfig{file=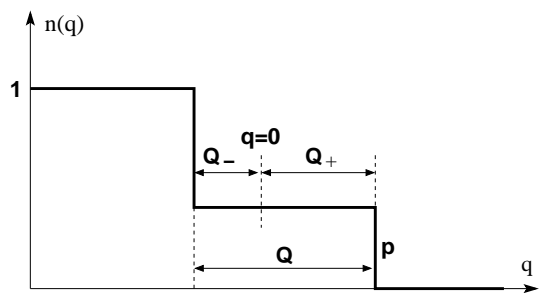,width=7cm}
\caption{Double step electron momentum distribution. The width $Q$ of the
step represents a voltage $\hbar vQ$ applied to a QPC, which has tunneling
probability $p$. The quantities $Q_+$ and $Q_-$ are determined following
Eq.~(\protect\ref{eqnqz}).}
\label{fig:1}
\end{figure}

We discuss in detail the long-time correlations for an initial state that
has a double-step momentum distribution as shown in the Fig.~\ref{fig:1},
which serves as a model for the electron distribution generated just after a
QPC in tunneling experiments.\cite{pierre, pierrenu2} For a QPC with a
tunneling probability $p$ at bias voltage $V$, steps in $\delta n(q)$ are
located at $q=-Q_{-}$ and $q=Q_{+}$, with separation $Q_{+}+Q_{-}\equiv
Q=eV/\hbar v_{F}$, so that 
\begin{equation*}
\delta n(q)=\left\{ 
\begin{array}{cc}
-(1-p) & \ \ \ -Q_{-}<q<0 \\ 
p & \qquad 0<q<Q_{+}%
\end{array}%
\right.
\end{equation*}%
and $\delta n(q)=0$ if $q$ lies outside this range. Eq.~(\ref{eqnqz})
implies %
$(1-p)Q_{-}=pQ_{+}$. %
With these definitions we have an effective temperature 
\begin{equation}
T^{\ast }=\sqrt{3}\left( \hbar v_{F}/\pi k_{B}\right) Q\sqrt{p(1-p)}\,.
\label{teffds}
\end{equation}

It is interesting to examine for this initial distribution how the
steady-state correlation function differs at long distances from a thermal
one with the same effective temperature. We have 
\begin{equation*}
\int_{-\infty }^{\infty }dq\ |\delta n(q)|^{2}=Qp(1-p),
\end{equation*}%
and 
\begin{equation*}
\int_{-\infty }^{\infty }dq\ \mathrm{sgn}(q)\ \delta n(q)=2Qp(1-p)
\end{equation*}%
so that for $Q|x|\gg 1$ 
\begin{equation}
f(x)=\pi p(1-p)Q|x|\,.  \label{f-dble}
\end{equation}%

For comparison, in a thermal state one has from Eq.~(\ref{freeGF}) 
\begin{equation}
f(x)=\frac{k_{B}T}{\hbar v_{F}}|x|\,.
\end{equation}%

A simple replacement of $T$ with $T^{\ast }$, expressed in terms of $p$ and $%
Q$ using Eq.~(\ref{teffds}), would lead one to expect in place of Eq.~(\ref%
{f-dble}) the result (also for $Q|x|\gg 1$) 
\begin{equation}
f(x)= \frac{\sqrt{3p(1-p)}}{\pi }Q|x|\,.
\end{equation}%
Clearly, (except for the special value $p(1-p)=3\pi ^{-4}$) these are
distinct forms. The difference is most pronounced if $p(1-p)\ll 1$, when the
correlation function in the steady state decays much more rapidly with $|x|$
than would be the case in a thermal state with the same energy density.

For this case of a double-step initial distribution, the steady-state
correlation function for general $x$ can be obtained explicitly, in the form 
\begin{equation}
\mathcal{G}\left( x\right) =\frac{i}{2\pi }\frac{1}{x+ia}\exp [-p(1-p)R(Qx)],
\label{gxlr}
\end{equation}%
where we have defined the function%
\begin{multline}
R(z)=\int_{0}^{1}\frac{dq}{q^{2}}(1-q)|e^{-iqz}-1|^{2}=  \label{IZ} \\
-2(1+\gamma_{E} +\ln z-\cos z-\mathrm{Ci}\,z-z\,\mathrm{Si}\,z),
\end{multline}
$\mathrm{Si}$ and $\mathrm{Ci}$ are the sine and cosine integral functions, and 
$\gamma_{E}\approx0.577216$ is the Euler constant. 

\subsubsection{Resulting long-time momentum distribution}

It is also interesting to discuss the steady-state momentum distribution
within this framework, especially in the regime $p(1-p) \ll 1$ where
deviations from thermal behaviour are largest. As a first step we use Eq.~(%
\ref{gxlr}) to derive a form for the real-space correlation function at
large $|x|$ that includes sub-leading terms omitted from Eq.~(\ref{FX}),
obtaining 
\begin{equation}
\mathcal{G}\left( x\right)\approx \frac{i}{2\pi }\frac{|e^{1+\gamma_{E}
}Qx|^{2p(1-p)}}{x+ia} e^{-\pi p(1-p)|Qx|}.
\end{equation}
From this, using Eq.~(\ref{mom_distr}), we find that the stationary momentum
distribution for $|k|\ll Q$ is 
\begin{equation}
n(k,t)=\frac{1}{2}-\frac{1}{\pi }e^{\alpha (1+\gamma_{E} )}\Gamma (\alpha )%
\mathrm{Im}[(-ik/Q+\pi \alpha /2)^{-\alpha }]\,,  \label{assimpt}
\end{equation}%
where $\alpha \equiv 2p(1-p).$ A striking point is that two scales appear in
Eq.~(\ref{assimpt}): one is $Q$, while the second one is $\pi \alpha Q/2$.
These scales are widely separated when $\alpha$ is small, and so there are
two distinct small $k$ regimes of behavior for $n(k,t)$ at long times, with 
\begin{equation}
n(k,t)= \frac{1}{2}-C ({\pi \alpha }/{2})^{-(1+\alpha )}\alpha k/Q
\end{equation}
for $k \ll \pi \alpha Q/2$ and 
\begin{equation}
n(k,t)= \frac{1}{2}-C ({\pi \alpha }/{2})[1-\alpha Q/k](Q/k)^{\alpha}
\end{equation}
for $\pi \alpha Q/2 \ll k \ll Q$, where $C =\frac{1}{\pi }e^{\alpha
(1+\gamma_E )}\Gamma (\alpha )$. 

\begin{figure}[bp]
\epsfig{file=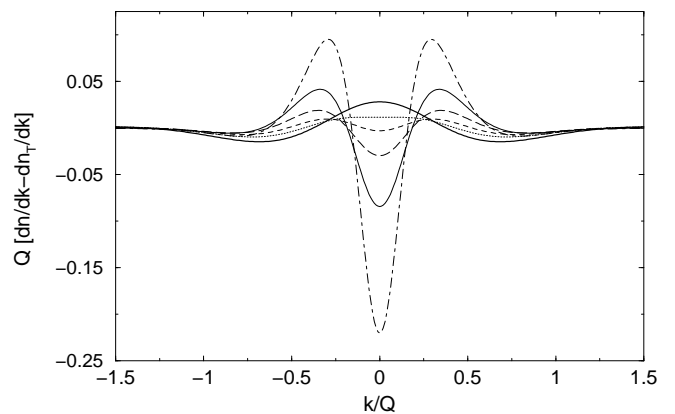,width=5.5cm,angle=-90}
\caption{Difference between derivatives of the electron momentum
distribution functions in an steady state described by Eq. (\protect\ref%
{gxlr}) and in a thermal state with the same energy density. Thick solid
line: $p=0.5$; dotted line: $p=0.3$; short-dashed line: $p=0.25$;
long-dashed line: $p=0.2$; thin solid line: $p=0.15$; dot-dashed line: $%
p=0.1 $.}
\label{mom_distr_deriv}
\end{figure}

As a supplement to this discussion of asymptotic forms, 
the full momentum distribution at long times can be obtained
numerically from Eqns.~(\ref{gxlr}) and (\ref{mom_distr}). We present its
behaviour in Fig. \ref{mom_distr_deriv}. Since the quantity measured
experimentally is a derivative, we plot the difference in derivatives
between the steady state distribution and the one for a thermal state with
the same energy density. As anticipated in our earlier discussion, these
differences are largest when $p(1-p)$ is small. %

\section{Equilibration at $\protect\nu =2$}

\label{sec:nu=2short}

In this section we study relaxation of a nonequilibrium distribution in a
system of two channels coupled by contact interactions. Whereas contact
interactions at $\nu=1$ simply renormalise the Fermi velocity, without
providing a mechanism for relaxation, at $\nu=2$ they give rise to two
linearly dispersing collective modes with distinct velocities, and hence 
\emph{do} produce relaxation. We therefore focus on this as the simplest 
interesting case.  We discuss briefly the effects of finite range interactions
at $\nu=2$ in Section \ref{nu2finite}.

The Hamiltonian of the system is given by Eq. (%
\ref{hamedge}), with $n_{c}=2$ and an interchannel interaction potential 
\begin{equation*}
U_{12}(x-x^{\prime })=U_{21}(x-x^{\prime })=(g/2)\delta (x-x^{\prime })\,.
\end{equation*}
Intrachannel interactions are absorbed into the Fermi velocities $v_{1,2}.$

We start this section with a general discussion of time evolution for this
Hamiltonian. Then we set out the solution to a special case, calculating the
electron correlation function at arbitrary time for a particular choice of
initial state, in which each channel is separately in thermal equilibrium,
but at different temperatures. Finally, we extend the approach described in
Section \ref{sec:nu=1} to calculate the long-times asymptotics of the
correlation function at long and short distances for an arbitrary initial
single-particle distribution.

\subsection{Diagonalization of bosonic Hamiltonian}

The bosonized form of the Hamiltonian is 
\begin{multline}
\hat{H}=\hbar v_{1}\sum_{k>0}k\hat{a}_{k}^{+}\hat{a}_{k}+\hbar
v_{2}\sum_{k>0}k\hat{b}_{k}^{+}\hat{b}_{k} \\
+\frac{1}{2}\frac{g}{2\pi }\sum_{k>0}k(\hat{a}_{k}\hat{b}_{k}^{+}+\hat{a}%
_{k}^{+}\hat{b}_{k}),  \label{07}
\end{multline}%
where $\hat{a}_{k}$ and $\hat{b}_{k}$ are the boson operators on the first
and the second channel. It can be diagonalized by the rotation 
\begin{equation}
(\hat{\alpha}_{k},\hat{\beta}_{k})^{T}=\mathrm{S}(\hat{a}_{k},\hat{b}%
_{k})^{T}\,,  \label{1}
\end{equation}%
with 
\begin{equation}
\mathrm{S}=\left( 
\begin{array}{cc}
\cos \theta  & \sin \theta  \\ 
-\sin \theta  & \cos \theta 
\end{array}%
\right) \,.
\end{equation}%

Setting $\gamma =g/(2\pi \hbar )$ and $\tan 2\theta =\gamma /(v_{1}-v_{2})$,
this brings the Hamiltonian to the diagonal form 
\begin{equation}
\hat{H}=\hbar v_{+}\sum_{k>0}k\hat{\alpha}_{k}^{+}\hat{\alpha}_{k}+\hbar
v_{-}\sum_{k>0}k\hat{\beta}_{k}^{+}\hat{\beta}_{k}\,.  \label{09}
\end{equation}%
The collective mode velocities $v_{\pm }$ are 
\begin{align}
v_{+}=& v_{1}\cos ^{2}\theta +v_{2}\sin ^{2}\theta +\tfrac{1}{2}\gamma \sin
2\theta ,  \notag \\
v_{-}=& v_{1}\sin ^{2}\theta +v_{2}\cos ^{2}\theta -\tfrac{1}{2}\gamma \sin
2\theta .  \notag
\end{align}%
The time dependence of the operators\ $\hat{\alpha _{k}}$ and $\hat{\beta
_{k}}$ in the Heisenberg representation is 
\begin{equation}
\hat{\alpha}_{k}(t)=e^{-iv_{+}kt}\hat{\alpha}_{k}\,,\quad \hat{\beta}%
_{k}(t)=e^{-iv_{-}kt}\hat{\beta}_{k}\,.  \label{3}
\end{equation}

Since we wish to discuss time evolution from an initial state specified by
occupation numbers for independent, non-interacting fermions, 
we need to re-express Eq. (\ref{3}) in terms of the operators $\hat{a}$ and $%
\hat{b}$ in their Schr\"{o}dinger representation, using the transformation $%
\mathrm{S}$ and its inverse. 
In this way we obtain 
\begin{align}
\hat{a}_{k}(t)& =\mathfrak{u}_{k}(t)\hat{a}_{k}+\mathfrak{s}_{k}(t)\hat{b}%
_{k}  \label{7} \\
\mathrm{and}\quad \hat{b}_{k}(t)& =\mathfrak{s}_{k}(t)\hat{a}_{k}+\mathfrak{v%
}_{k}(t)\hat{b}_{k}.  \notag
\end{align}%
with coefficients 
\begin{align}
\mathfrak{u}_{k}(t)& =e^{-iv_{+}kt}\cos ^{2}\theta +e^{-iv_{-}kt}\sin
^{2}\theta ,\   \label{8} \\
\mathfrak{v}_{k}(t)& =e^{-iv_{+}kt}\sin ^{2}\theta +e^{-iv_{-}kt}\cos
^{2}\theta ,\   \notag \\
\mathrm{and}\quad \mathfrak{s}_{k}(t)& =\left(
e^{-iv_{+}kt}-e^{-iv_{-}kt}\right) \cos \theta \sin \theta .  \notag
\end{align}%
Equations (\ref{7}) and (\ref{8}) define a unitary transformation of the
boson operators which generates their time evolution independently for each
wavevector.

In bosonized form, the electron correlation function is 
\begin{multline}
G_{\eta }(x,t)=\langle e^{i\hat{H}t}\hat{\psi}_{\eta }^{+}(x)\hat{\psi}%
_{\eta }(0)e^{-i\hat{H}t}\rangle \\
=(2\pi a)^{-1}\langle e^{i\hat{\phi}_{\eta }(x,t)}e^{-i\hat{\phi}_{\eta
}(0,t)}\rangle\,.  \label{12}
\end{multline}%
For an initial state without correlations between the channels, this
factorises into the product 
\begin{equation}
G_{\eta }(x,t)=(2\pi a)^{-1}G_{a}^{(\eta )}(x,t)G_{b}^{(\eta )}(x,t),
\label{16}
\end{equation}%
where 
\begin{equation*}
G_{a,b}^{(\eta )}(x,t)=\langle e^{i\hat{\phi}_{a,b}^{(\eta )}(x,t)}e^{-i\phi
_{a,b}^{(\eta )}(0,t)}\rangle
\end{equation*}%
and [from Eq. (\ref{7})] %
\begin{equation}
\hat{\phi}_{a}^{(1)}(x,t)=-\sum_{q>0}(2\pi /qL)^{1/2}(\mathfrak{u}_{q}(t)%
\hat{a}_{q}e^{iqx}+\mathrm{h.c.})e^{-qa/2}\,.  \label{phib1}
\end{equation}%
The corresponding expression for $\hat{\phi}_{b}^{(1)}\left( x,t\right) $ is
obtained by making the substitutions $\mathfrak{u}_{q}(t)\rightarrow 
\mathfrak{s}_{q}(t)$ and $\hat{a}_{q}\rightarrow \hat{b}_{q}$ in Eq. (\ref%
{phib1}). Similarly, we obtain an expression for bosonic fields in the
second channel by making the substitutions $\mathfrak{s}_{q}\rightarrow 
\mathfrak{v}_{q}$ and $\mathfrak{u}_{q}\rightarrow \mathfrak{s}_{q}$ in the
equations for $\hat{\phi}_{a,b}^{(1)}(x,t)$.

\subsection{Equilibration of channels with unequal initial temperatures}
\label{different-temps}

We next study relaxation from an initial state in which the particles in
channels of the system without interactions have thermal distributions with
different temperatures, $T_{\eta }$. A similar problem was studied in a
context of a quantum quench for a non-chiral Luttinger liquid, in Ref.~%
\onlinecite{cazalilla}.

The simplifying feature of a thermal initial state is that there is
uncorrelated occupation, both of the fermion orbitals on which $\hat{c}%
_{k\eta }$ act, and of the boson orbitals on which $\hat{a}_{k}$ and $\hat{b}%
_{k}$ act. We can therefore specify this state by the averages 
\begin{equation}
\langle \hat{a}_{k}^{+}\hat{a}_{k}\rangle =(e^{\hbar v_{1}k\beta
_{1}}-1)^{-1},\ \ \langle \hat{b}_{k}^{+}\hat{b}_{k}\rangle =(e^{\hbar
v_{2}k\beta _{2}}-1)^{-1},  \label{9}
\end{equation}%
where $\beta _{\eta }=1/k_{B}T_{\eta }$. Correlators of the fields $\hat{\phi%
}_{a,b}^{(\eta )}(x,t)$ are those for a non-interacting edge, with the
appropriate choice for velocity. We use the notation $w_{\eta }=-2i\beta
_{\eta }\hbar v_{\eta }/L$ and $\xi _{\eta }=\pi /\beta _{\eta }\hbar
v_{\eta }$. We also define $r=\frac{1}{2}\sin ^{2}2\theta $, $s=1-r$, and 
\begin{equation}
v_{D}=v_{+}-v_{-}=\dfrac{\gamma }{\sin 2\theta }\,.
\end{equation}%
In this way we obtain 
\begin{multline}
G_{a}^{(1)}(x,t)=(2\pi a/L)^{s}\frac{1}{w_{1}^{s}}\frac{1}{\sinh ^{s}[\xi
_{1}(x+ia)]}\times \\
\left[ \frac{\sinh [\xi _{1}(v_{D}t+ia)]\sinh [\xi _{1}(-v_{D}t+ia)]}{\sinh
[\xi _{1}(x+v_{D}t+ia)]\sinh [\xi _{1}(x-v_{D}t+ia)]}\right] ^{r/2}
\label{36}
\end{multline}%
In a similar way we get for $G_{b}^{(1)}(x,t)$
\begin{multline}
G_{b}^{(1)}(x,t)=(2\pi a/L)^{r}\frac{1}{w_{2}^{r}}\frac{1}{\sinh ^{r}[\xi
_{2}(x+ia)]}\times \\
\left[ \frac{\sinh [\xi _{2}(x+v_{D}t+ia)]\sinh [\xi _{2}(x-v_{D}t+ia)]}{%
\sinh [\xi _{2}(v_{D}t+ia)]\sinh [\xi _{2}(-v_{D}t+ia)]}\right] ^{r/2}
\label{39}
\end{multline}%

As a check, note that the initial electron correlation function 
\begin{equation*}
\langle \hat{G}_{\eta }(x,0)\rangle =\left( i/2\pi \right) \xi _{\eta }\sinh
^{-1}[\xi _{\eta }(x+ia)]
\end{equation*}%
derived in this way from Eq.~(\ref{16}) coincides with the thermal one at
inverse temperature $\beta _{\eta }$. Next consider the long-time
asymptotics of $G_{\eta }(x,t).$ The function $1/\sinh [\xi _{\eta }(x+ia)]$
is peaked near $x=0$ and falls off exponentially at large $x$ on a scale $%
1/\xi _{\eta }$ (or as a power-law at zero temperature). For a fixed $x$ at
long times, the factor $[\ldots ]^{r/2}\ $approaches one. The lengthscale
defined by the product $\sinh [\xi _{1}(x+ia)]^{-s}\sinh [\xi
_{2}(x+ia)]^{-r}$ is 
\begin{equation*}
l=(s\xi _{1}+r\xi _{2})^{-1}=\frac{\hbar }{\pi }\left[ \frac{s}{\beta
_{1}v_{1}}+\frac{r}{\beta _{2}v_{2}}\right] ^{-1}.
\end{equation*}%
The time taken to approach the asymptotic form is $\tau \approx l/v_{D}$.
For times much longer than $\tau $ we have 
\begin{multline}  \label{eqtwochbeta}
G_1(x)=\frac{i}{2\hbar }\frac{1}{\left( \beta _{1}v_{1}\right) ^{s}}\frac{1}{%
\sinh ^{s}[\xi _{1}(x+ia)]} \\
\times \frac{1}{\left( \beta _{2}v_{2}\right) ^{r}}\frac{1}{\sinh ^{r}[\xi
_{2}(x+ia)]}.
\end{multline}

\begin{figure}[bp]
\epsfig{file=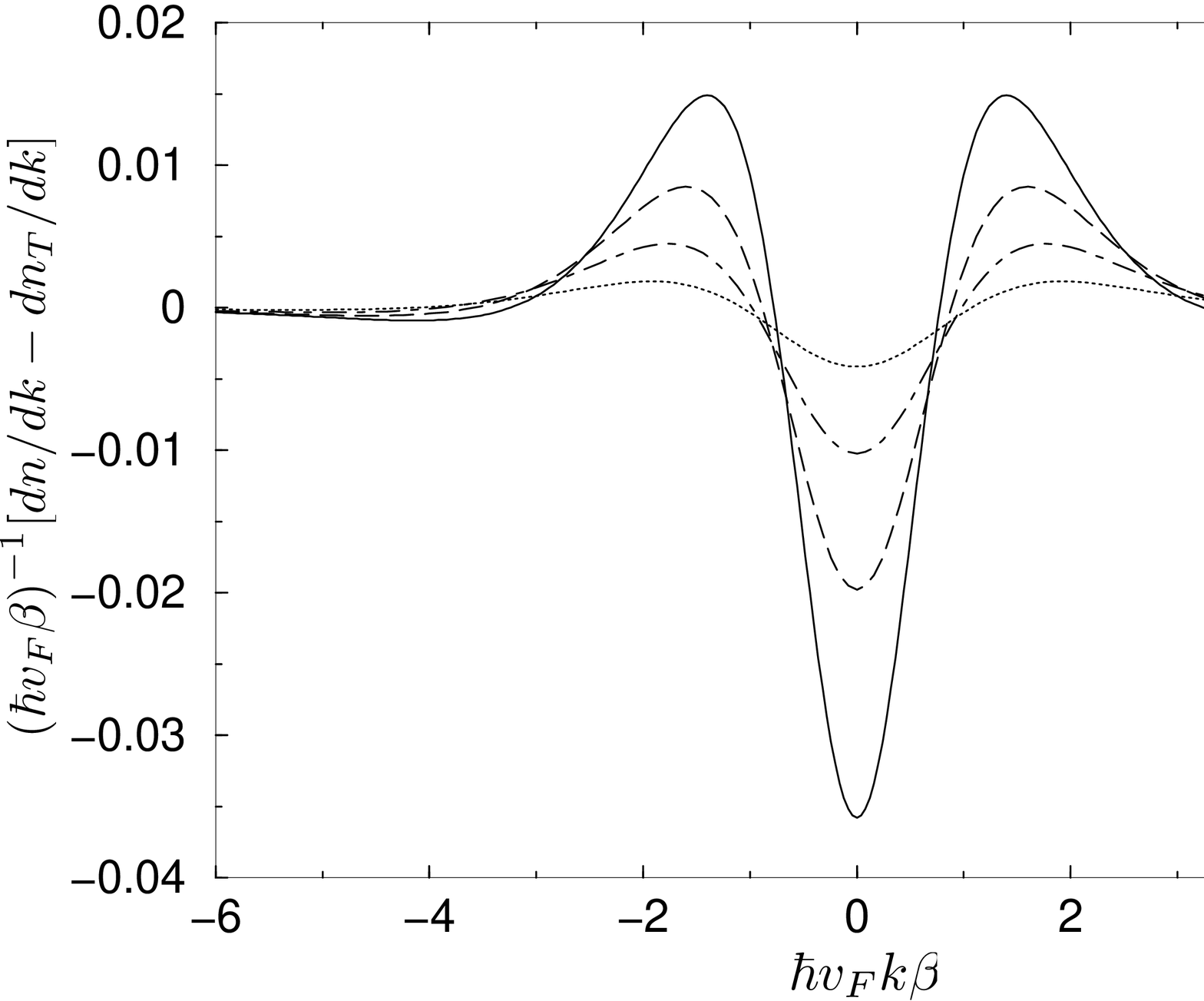,width=8cm,angle=0}
\caption{Difference between derivatives of the electron momentum
distribution functions in an steady state described by Eq. (\protect\ref%
{eqtwochbeta}) with $T^{*}_2=0$ and in a thermal state at temperature $T=%
\protect\sqrt{s}T^{*}_1$. Thick solid line: $\protect\theta=\protect\pi/4$;
dashed line: $\protect\theta=\protect\pi/6$; dot-dashed line: $\protect\theta%
=\protect\pi/8$; dotted line: $\protect\theta=\protect\pi/12$.}
\label{mom_distr_deriv_exact}
\end{figure}

We see that at long times the system reaches a steady state. In this state
the electron correlation function is strikingly different from a thermal
one, being given by a product of two thermal correlation functions for
systems at different temperatures, each raised to powers that depend on the
strength of the interchannel interactions. Comparison of electron momentum
distribution obtained from Eq. (\ref{eqtwochbeta}) with a thermal one is
presented in Fig. \ref{mom_distr_deriv_exact}.

\begin{figure}[bp]
\epsfig{file=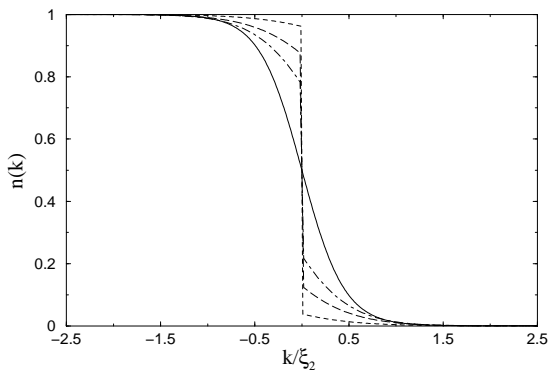,width=5cm,angle=-90}
\caption{Time evolution of the electron momentum distribution for a system of two channels in 
the strong-couping limit $(\theta=\pi/4)$ with unequal initial temperatures
$T^{*}_1=0$ and $T^{*}_2$, obtained using Eqns.~(59), (63), and (64). Dotted line: $\xi_2 v_D t=1$;
dashed line: $\xi_2 v_D t=2$; dot-dashed line: $\xi_2 v_D t=3$. Solid line: thermal distribution
at temperature $T=T^{*}_2/\sqrt{2}$, shown for comparison.}
\label{fig5}
\end{figure}

Clearly, an attraction of studying evolution from this special class of initial states
is that one obtains explicit results at not only at long times but at all intermediate times. 
In Fig.~\ref{fig5} we exploit this to show the time evolution of the momentum distribution 
in a channel initially at zero temperature, coupled to a second channel with non-zero initial temperature.

\subsection{Correlations at $\protect\nu{=}2$ in terms of initial density
operator}

As a second example of relaxation in the two-channel system, we consider an
initial state with a double step distribution of electron momentum in one
channel. The correlation function decouples into a product [Eq. (\ref{16})]
of factors that can each be expressed using the approach of Section~\ref%
{correlations-via-density} in terms of the initial electron density
operator. For example, 
\begin{equation}
G_{a}^{(1)}(x,t)=e^{\frac{1}{2}[\hat{\phi}_{a}(x,t),\hat{\phi}_{a}(x^{\prime
},t)]}\langle e^{i\int K(x,x^{\prime },t;y)\hat{\rho}(y)dy}\rangle ,
\label{GaK}
\end{equation}%
where the kernel is given by 
\begin{equation}
K(x,x^{\prime },t;y)=-i\sum_{q\neq 0}(2\pi /qL)\mathfrak{u}_{k}^{\ast
}(t)(e^{-iqx}-e^{-iqx^{\prime }})e^{iqy-|q|a}
\end{equation}%
and has the form 
\begin{multline*}
K(x,x^{\prime },t;y)=\pi \big(\big[\mathrm{sgn}(v_{+}t-x+y) \\
-\mathrm{sgn}(v_{+}t-x^{\prime }+y)\big]\cos ^{2}\theta \\
+\big[\mathrm{sgn}(v_{-}t-x+y)-\mathrm{sgn}(v_{-}t-x^{\prime }+y)\big]\sin
^{2}\theta \big)\,.
\end{multline*}%
At times $t>|(x-x^{\prime })/v_{D}|$ this kernel separates into two steps:
one occupying the interval $(x^{\prime }-v_{+}t,x-v_{+}t)$, and the other in 
$(x^{\prime }-v_{-}t,x-v_{-}t)$. The separation between these intervals
grows at speed $v_{D}t$. An energy scale $eV$ in the initial state sets a 
characteristic length $\hbar v/eV$: the timescale $\tau$ 
introduced in Sec. II is the time taken for the separation between steps to reach this characteristic length.
Introducing an operator 
\begin{equation*}
\hat{N}(x)=\int_{0}^{x}\hat{\rho}(z)dz
\end{equation*}%
that counts the number of particles in an interval of length $x$, in the
long time limit there is the factorization%
\begin{equation*}
\langle e^{i\int K(x,0,t;y)\hat{\rho}(y)dy}\rangle =\langle e^{-2i\pi \hat{N}%
(x)\cos ^{2}\theta }\rangle \langle e^{-2i\pi \hat{N}(x)\sin ^{2}\theta
}\rangle .
\end{equation*}

Here we see a contrast between our earlier results at $\nu=1$ in systems
with finite range or Coulomb interactions, for which $\omega_q$ has
dispersion, and these results at $\nu=2$ with contact interactions between
electrons in different channels, for which the bosonic modes have linear
dispersion relations. While in the former case the steady-state correlation
function can be expressed solely in terms of the two-point density
correlation function of the initial state [see Eq.~(\ref{GLT2})], this is
not so in the latter case. Nevertheless, simplifications arise in the small
and large $x$ asymptotics of the correlation function, as we now show.

\subsubsection{Short-distance asymptotics}

The short-distance form of the steady-state correlation function is fixed by
the energy density in each channel in the initial state, characterised by
effective temperatures $T_{\eta }^{\ast }$. Setting $\xi _{\eta }^{\ast
}=\pi k_{B}T_{\eta }^{\ast }/\hbar v_{\eta }$, we find%
\begin{equation*}
G_1\left( x\right) =\frac{i}{2\pi }\frac{1}{x+ia}\big[{1+\frac{x^{2}}{3!}%
(s\xi_{1}^{*2}+r\xi _{2}^{*2})+\ldots }\big]^{-1}\,.
\end{equation*}%
Correlations at short distances in channel 1 therefore have an effective
temperature 
\begin{equation}
T_{\mathrm{eff}}^{2}=sT_{1}^{\ast 2}+rT_{2}^{\ast 2}\,,  \label{effT1}
\end{equation}%
with the corresponding result for channel 2 obtained by exchanging $r$ and $%
s $. We see that the effective temperatures of the two channels are
different, except in the strong-coupling limit where $r=s=1/2$.

\subsubsection{Large-distance asymptotics}

Calculation of the large $x$ asymptotics of the Green function in the
form given by exponent of the electron counting operator is a well-studied
problem. The asymptotics can be obtained using the theory of Toeplitz determinants:
see for example Ref.\onlinecite{mirlin}. 

In the case of a double-step distribution of
width $Q$, which is created and measured in channel 1, with channel 2 initially at zero
temperature, we obtain for $Q|x|\gg 1$ in the leading order
\begin{equation}
G_1(x)\approx \frac{i}{2\pi }\frac{1}{x+ia}\exp [-\bar{s}\frac{|Qx|}{2\pi}],
\label{twoch2}
\end{equation}
where
\begin{multline}\nonumber
\bar{s}=-\frac{1}{2}[\ln(1-4p(1-p)\sin^2[\pi\cos^{2}\theta])\\+\ln(1-4p(1-p)\sin^2[\pi\sin^{2}\theta])];
\end{multline}
see Ref.\onlinecite{mirlin,exact2} for details.

\subsubsection{Finite range interactions at $\nu=2$}\label{nu2finite}

Finite range interactions, in contrast to contact interactions, generate dispersion
so that the mode velocities $v_\pm$ and the mixing angle $\theta$ become functions
of wavevector. Dispersion in the two mode velocities provides a further mechanism for
relaxation at $\nu =2$. One sees, however, from Eqns.~(1) and (2) that 
this additional mechanism is less important at small $V$ than the one 
due to contact interactions, since it yields a much longer relaxation time (scaling as $V^{-3}$
in place of $V^{-1}$). Thus at small $V$ there is window between these two time scales
during which the results for contact interactions apply even in the presence
of dispersion.  At times beyond the longer scale, our results
for $\nu=1$ carry over with simple modifications and we obtain for $t \to \infty$
\begin{equation}
 G_{1}(x)= [{\cal G}_a(x)]^{\cos^4\theta + \sin^4 \theta} [{\cal G}_b(x)]^{2\cos^2 \theta \sin^2 \theta}
\end{equation}
Here ${\cal G}_a(x)$ and ${\cal G}_b(x)$ are obtained from the expression for
${\cal G}(x)$  in Eq.~(\ref{GLT2}) by computing expectations values in the initial stares of channel
$1$ and $2$ respectively, while $\theta$ denotes the limiting
value of the mixing angle at small vectors.

\section{Tunneling density of states}

\label{different-t}

As summarised in Section \ref{sec:discussion}, the experiments of Refs.~\onlinecite{pierre,pierrenu2,altimiras} probe the electron distribution by means of tunneling measurements. With that as a motivation, we study in this section the 
tunneling density of states (TDOS). We consider in particular edge channels at $\nu=2$ after 
long-time evolution from an initial state 
characterised by two temperatures, as introduced in Section \ref{different-temps}.
We focus on this case, rather than an initial state with a double-step distribution
which would better represent the experimental situation, because the analytic results 
available for the Green function with these initial conditions facilitate accurate
evaluation of the TDOS. 
It is important to stress that, while for a non-interacting system the electron
momentum distribution and the tunneling density of states are simply related, 
with interactions they become independent quantities. This is illustrated (in the space
and time, rather than momentum and energy domains) in our results below by contrasting
Eq.~(\ref{eqtwochbeta}) with Eq.~(\ref{tdos}).

For a single edge state in equilibrium at temperature $T$ with only contact interactions,
the TDOS defined in Eq. (\ref{DOS1}) has the form
\begin{equation}
\nu (E)=(\hbar v)^{-1}n_{F}(E/k_{B}T)\,,  \label{nuEeq}
\end{equation}
and so the tunneling current is proportional to the Fermi distribution 
$n_F(x) = (e^x+1)^{-1}$.
More generally, one can attempt to characterise the energy dependence of the TDOS by an effective temperature $T_{\rm eff}$. As there is no unique way to do so, we explore two alternative. First, since the energy density in thermal equilibrium is $\frac{\pi}{12}(k_{\mathrm B}T)^2$, one natural definition of  $T_{\rm eff}$ is via
\begin{equation}
\frac{\pi }{12}(k_{\mathrm{B}}T_{\mathrm{eff}})^{2}=(2\pi \nu _{\mathrm{eff}%
})^{-1}\int_{-\infty }^{\infty }[\nu (E)-\nu_{\mathrm{eff}}\theta (-E)]EdE\,,  \label{ebar}
\end{equation}
where $\theta(E)$ is the Heaviside step function, and the normalisation is $\nu_{\mathrm{eff}} \equiv \nu(-\infty)$, the TDOS far from the Fermi energy.
To evaluate $\nu(E)$ we require [see Eq.~(\ref{DOS1})]
the correlator
\begin{equation*}
G_{\eta }(\tau)=\lim_{t\rightarrow \infty }\langle \hat{\psi}^{+}(x,t+\tau
)\psi (x,t)\rangle.
\end{equation*}

Consider the two-channel problem discussed in Section \ref{sec:nu=2short}. In
thermal equilibrium the correlator $G_{1}(t)$ reads%
\begin{multline*}
G_{1}(t)=\frac{i}{2\beta v_{+}^{\cos ^{2}\theta }v_{-}^{\sin ^{2}\theta }}%
\frac{1}{\sinh [\tfrac{\pi }{\beta v_{+}}(-v_{+}t+ia)]^{\cos ^{2}\theta }} \\
\times \frac{1}{\sinh [\tfrac{\pi }{\beta v_{-}}(-v_{-}t+ia)]^{\sin
^{2}\theta }},
\end{multline*}%
which gives for the density of states%
\begin{equation*}
\nu (E)=(\hbar v_{+}^{\cos ^{2}\theta }v_{-}^{\sin ^{2}\theta
})^{-1}n_{F}(E/k_{B}T)\text{.}
\end{equation*}%
Thus the equilibrium TDOS in a two channel system with contact
interactions between electrons in each channel is proportional to the Fermi distribution and independent of interaction strength.
It is worth noting that this is not the case for a single channel with finite-range or
Coulomb interactions. 
\begin{figure}[tp]
\epsfig{file=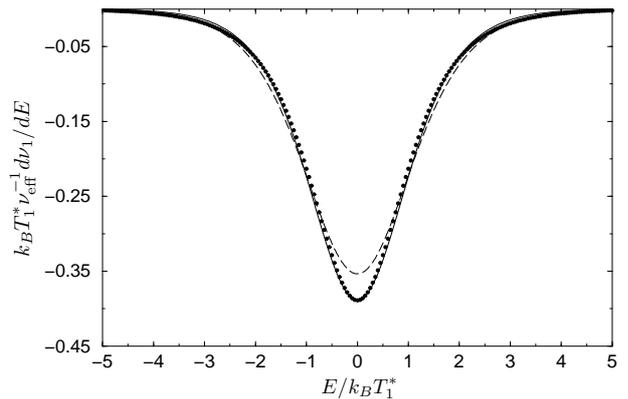,width=8cm,angle=0}
\caption{Derivative of the normalized tunneling density of states as a
function of energy in a two-channel system. Filled circles: in the steady
state reached with initial temperatures $T_{1}^{\ast }>0$ and $T_{2}^{\ast }=0$
at $g/\hbar v_{1}=1$ in the strong coupling limit $v_1=v_2$. 
Dashed line: in a thermal state at the temperature $T_{1}^{\ast }/\protect%
\sqrt{2}$ expected from equipartition. Solid line: in a thermal state with a
temperature $0.64\times T_{1}^{\ast }$ chosen to give the best fit to the
filled circles.}
\label{fig:dnde2ch}
\end{figure}

Now let us consider the tunneling density of states at long times, when
the system starts out of equilibrium. In order to have explicit results
that are easy to evaluate accurately for all energies, we consider an 
initial state with channels at unequal temperatures $T_{1,2}^{\ast }.$ 
For this case the Green function in channel
\textquotedblleft $1$\textquotedblright\ at long times reads%
\begin{eqnarray}\label{tdos}
G_{1}(t)&=&L^{-1}w_{1}^{-s}w_{2}^{-r}\times \nonumber \\
&\times& \frac{1}{\sinh [\xi _{2}(-v_{+}t+ia)]^{r/2}\sinh [\xi
_{2}(-v_{-}t+ia)]^{r/2}} \nonumber \\
&\times& \frac{1}{\sinh [\xi _{1}(-v_{+}t+ia)]^{\cos ^{4}\theta }\sinh [\xi
_{1}(-v_{-}t+ia)]^{\sin ^{4}\theta }}\nonumber.\\
\end{eqnarray}%
This gives for the integral Eq. (\ref{ebar})%
\begin{multline}
\frac{\pi }{12}(k_{\mathrm{B}}T_{\mathrm{eff}})^{2}=\frac{\pi (k_{\mathrm{B}%
}T_{1}^{\ast })^{2}}{24v_{1}^{2}}[(v_{+}^{2}-v_{-}^{2})\cos 2\theta +
\label{Eexc} \\
+(v_{+}^{2}+v_{-}^{2})\{s+(v_{1}T^{*}_{2}/v_{2}T^{*}_{1})^{2}r\}]\,.
\end{multline}%
The effective temperature $T_{\mathrm{eff}}$ obtained from Eq. (\ref{Eexc}) is a
function of two parameters: $\theta $ and $v_{2}/v_{1}$. For $v_{1}=v_{2}$
and with the second channel at an initial temperature $T_{2}^{\ast }=0$ we obtain%
\begin{equation}
\frac{\pi }{12}(k_{\mathrm{B}}T_{\mathrm{eff}})^{2}=\frac{\pi }{24}%
(k_{B}T_{1}^{\ast })^{2}[1+(\gamma /2v_{1})^{2}].  \label{eqenergy2}
\end{equation}%
From this we find $T_{\mathrm{eff}}=(T_{1}^{\ast }/%
\sqrt{2})[1+(\gamma /2v_{1})^{2}]^{1/2},$ which is always higher than the
equipartition value $T_{1}^{\ast }/\sqrt{2}$. 

An alternative way to define $T_{\mathrm{eff}}$ is to fit the
normalized TDOS to a thermal Fermi function, as displayed in Fig. \ref%
{fig:dnde2ch}. 
Interestingly, with $v_1=v_2$ and for $g/\hbar v_1\lesssim 1$,
the best fit 
yields $T_{\mathrm{eff}}\sim 0.64\times T_{1}^{\ast },$ which is about $9\%$ lower than
the equipartition result. 

We believe these calculations of effective temperature make some useful points. First,
they show that there is no unique definition of effective temperature because the steady state is
not thermal. Second, they serve to demonstrate that  it is possible for effective temperature (at least,
by our second definition and for the initial state we have treated) to be lower than expected from
equipartition. Further, the specific mechanism for a low effective temperature is clear: the 
derivative of the tunneling density of states at long times has less weight in its peak
and more weight in its flanks than in thermal equilibrium. 

In the experiments of Refs.~\onlinecite{pierre} and \onlinecite{pierrenu2} an apparent energy loss is reported
from analysis of the TDOS far from the QPC, compared to behaviour close to the QPC. 
The experimental analysis uses two methods to find an effective temperature, corresponding to
the two definitions we have employed. Although no systematic discrepancies are reported 
between results from the two methods, the first method (calculation of
energy density by integration) relies on capturing contributions from the flanks of the
distribution, which in turn requires an accurate value for the baseline. It may therefore
be less robust than the second method (fitting to a Fermi function).
We find it interesting to note that the temperature reduction we obtain from this method is within the errorbars
of the found in Ref.~\onlinecite{pierrenu2}.

\section{Acknowledgements}
We acknowledge discussions with L.~Glazman, Y.~Gefen, A.~Mirlin, D.~Mailly and F.~Pierre.
This work was supported by EPSRC under grant EP/D050952/1.


\begin{thebibliography}{99}
\bibitem{footnote} Using the notation of Ref.~\onlinecite{degiovanni}, the quantity we denote by $\tau$
in Eq.~(\ref{tau}) is $L_{V}/v_0$.
\end{thebibliography}
\end{document}